\documentclass[prc,twocolumn,showpacs,preprintnumbers,amsmath,amssymb]{revtex4}
\usepackage{graphicx}
\usepackage{dcolumn}
\usepackage{bm}
\input{epsf.tex}

\def\be{\begin{equation}}
\def\ee{\end{equation}}
\def\ba{\begin{array}}
\def\ea{\end{array}}
\def\bea{\begin{eqnarray}}
\def\eea{\end{eqnarray}}
\begin{document}

\title{Evidence of a proton halo in $^{23}$Al: A mean field analysis}

\author{\large $^{1}$R. N. Panda}
\thanks{rnpanda@iopb.res.in}
\author{\large $^{2}$M. Panigrahi}
\affiliation{$^{1,2}$Department of Physics, Siksha O Anusandhan University, Bhubaneswar-751 030, India}
\author{\large $^{3}$Mahesh K. Sharma}
\affiliation{$^3$Department of Physics, A. S. College, Khanna, Ludhiana, Punjab 141 417, India}
\author{\large $^{4}$S. K. Patra}
\affiliation{$^4$Institute of Physics, Sahcivalaya Marg Bhubaneswar-751 005, India}
\affiliation{$^4$Homi Bhaba National Institute, Training School Complex, Anushakti Nagar, Mumbai 400 085, India}

\date{\today}

\begin{abstract}
We have studied the nuclear structure properties like binding energy, charge radius, quadrupole
deformation parameter for various isotopes of Al from the valley of stability
to drip line region using the well known relativistic mean field formalism (RMF) with NL3 parameter set. We have compared our results with experimental data and found reasonable agreement. Further, we have taken spherical
and deformed RMF densities to estimate the reaction dynamics of  $^{23-28}$Al
isotopes as projectiles and $^{12}$C as target by conjunction in Glauber
model. The estimated results are also compared with the experimental data. The analysis of angular elastic differential and one proton removal cross sections are also studied with Glauber model many body system to investigate the structural feature of $^{23}$Al. The evidence of enhanced
total reaction cross section, higher value of rms radius, narrow longitudinal momentum distribution and small proton separation energy of $^{23}$Al support its proton halo structure.
\end{abstract}


\maketitle

\section{\label{sec:level1} INTRODUCTION}
In past few decades, considerable theoretical interest has focused on the
studies of exotic nuclei since the first observation on existence of the
neutron halo in $^{11}$Li \cite{tanihata1985}. Further the existence of neutron
halo in other neutron-rich nuclei \cite{jensen2001} such as $^{11,14}$Be,
$^{17}$B, $^{19,22}$C \textit{etc}. enhanced the search for such candidates in other
region of mass table. However, on the proton halo, the number of experiments
are relatively few, as identification of a proton halo is more difficult due to Coulomb barrier. Now a days, there is evidence for the proton halo in very
proton-rich nuclei like $^8$B, $^{17}$Ne and $^{26,27,28}$P
\cite{minamisono1992,cortina2003,kanungo2003,patra1998}. Borcea et al.
\cite{borcea1997} systematically studied about $^{8}$B and pointed out a pigmy
halo in it. The evidence of a proton halo in the 1st excited state of $^{17}$F
is suggested by Morlock \textit{et al.} \cite{morlock1997}. The structure of
$^{23}$Al, a possible candidate for one-proton halo has created interest for
 both theorists as well as experimentalists. Large size of $^{23}$Al may be due to its proton halo structure or the existence of a large size $^{22}$Mg core
\cite{cai2002,ravinder2009,ozawa2006,ozawa1996,zhang2002}. Thus, the existence
of proton halo in $^{23}$Al fill up the gap between $^{17}$Ne and
$^{26,27,28}$P in the context of halo phenomena \cite{ren2003}.
This type of study become feasible due to the development of Radioactive Ion Beams (RIBs) techniques in various laboratories around the globe.
Experimental techniques and theoretical analysis are commonly used to
collect many important information about the structure, such as nuclear size, valence nucleon
distribution and halo structure of exotic nuclei. The investigation of
various reaction variables like total reaction cross-section $\sigma_R$,
one- and two- nucleon removal cross-section ($\sigma_{-1N}$, $\sigma_{-2N}$)
and the longitudinal momentum distribution $P_{||}$ are some of
the required quantities for such studies.
Total reaction cross section, $\sigma_R$ and nucleon removal cross section,
$\sigma_{-N}$ provides useful information regarding matter distribution
and the size of atomic nuclei \cite{skpatra1,panda2011,panda2014}.
Reaction cross section $\sigma_R$ mainly describes the rms nuclear radius, however, $\sigma_{-N}$ is sensitive only to the surface distribution. Thus combined information for both $\sigma_R$ and
$\sigma_{-N}$ are necessary in nuclear matter distribution study \cite{banu2011}. $\sigma_R$ and $\sigma_{-N}$ can be used to predict proton halo nuclei.

Recently, many theoretical models have been developed on the study of nuclear
reactions involving weakly bound nuclei. Relativistic mean field (RMF) theory
has been very useful to study elastic and breakup observables for these nuclei
\cite{patra91,serot86,pring96,lala97,panda2009}.
The authors of Refs: \cite{cai2002,zhang2002} studied the configuration of a
core nucleus ($^{22}$Mg) plus a valence proton using Glauber model \cite{gla59}. They predict that $^{23}$Al should have a long tail in its density distribution which is one of
the characteristic feature of a halo nucleus. Some other authors
\cite{ren1996,chen1998}, using RMF model, predicted one-proton halos in
$^{26,27,28}$P and two proton
halo in $^{27,29}$S. Wang et al. \cite{wang1997} predicted the halo structure
of $^{23}$Al by analysing rms radii of nuclei in RMF calculations.
It is conventional that the large spatial distribution of the valence  nucleon is reflected by narrow fragment momentum distribution  and also there is a correlation between
the magnitude of $\sigma_{-1p}$ with
the width of the $P_{||}$, while moving toward the nucleon drip-lines. In addition to this, one-proton removal reaction cross-sections gives some important information on nuclear structure which supports the results obtained  from $P_{||}$.
The very small binding of the valence particles is one of the important reason for the formation of halo nucleus near the drip-line . The quadrupole
deformation of the halo is determined by the structure of the weakly
bound valence orbital and is independent of the shape of the core
\cite{misu97}. Further, in a deformed nucleus, the existence of two nucleon halo is a rare phenomenon \cite{nunes05}. Nunes \cite{nunes05} calculations with a variety of 3-body NN tensor force which goes
beyond the unusual pairing in Hartree-Fock-Bogoliubov (HFB) and the coupling
due to core deformation/polarization significantly reduces the formation of
3-body Borromean systems. In a relatively recent work \cite{zhou10} halo phenomena
in deformed nuclei are explained within deformed Relativistic
Hartree Bogoliubov (RHB) theory and their finding
in weakly bound $^{44}$Mg  nucleus indicates a decoupling of the halo orbital
from the deformed core, which is in  agreement with the conclusion of Ref. \cite{misu97}.\\
In the present work, our interest is to study the bulk properties like
binding energy (BE), charge radius ($r_{ch}$), quadrupole deformation parameter
($\beta_2$), proton and neutron separation energy ($S_p, S_n$) $etc$. from stable nuclei to
drip line region and to investigate the reaction cross-sections, $\sigma_R$ for $^{23-28}$Al isotopes as projectiles and $^{12}$C
as target in the framework of Glauber formalism \cite{gla59}, using densities from well-known relativistic mean field (RMF) model with NL3 parameter set
 \cite{skpatra1,panda2009,sharma06}. The angular elastic differential, one proton removal cross
section and angular momentum distribution are also discussed in order to study
the structural feature of $^{23}$Al.\\

The paper is organised as follows:
The RMF formalism and the reaction mechanism in the framework of
Glauber model are explained briefly in section II. The results obtained
from our calculations are discussed in section III. In this section,
we mean to study the relevance of Glauber model in the
context of both stable and drip-line nuclei particularly those with halo
structure. Finally, a brief summary and concluding
remarks are given in the last section IV.\\
\section{Theoretical Framework}
\subsection{RMF Lagrangian density}
The successful applications of RMF both in finite and infinite
nuclear systems make more popular of the formalism in the present decades.
The RMF model has been extended to the Relativistic Hartree-Bogoliubov (RHB)
and density functional approach to study the static and dynamic aspects of exotic nuclear structure
\cite{vretenar05,meng06}.
The use of RMF formalism for finite nuclei as well as infinite
nuclear matter are well documented and details can be found in Refs:
\cite{patra91,serot86,pring96,lala97} and
\cite{tang97,patra01,skpatra01,serot97} respectively.
The working expressions for the density profile and other related quantities are
available in
\cite{skpatra1,patra91,serot86,pring96,panda2009,sharma06,tang97,patra01,skpatra01,serot97}. For completeness, the RMF Lagrangian is written as
\begin{eqnarray}
{\cal L}
&=& \overline{\psi}_i(i\gamma^{\mu}\partial_{\mu}-M)\psi_i+{\frac{1}{2}}\partial^{\mu}\sigma\partial_{\mu}\sigma\nonumber\\
&&-\frac{1}{2}m^{2}_{\sigma}\sigma^{2}-\frac{1}{3}g_2\sigma^3-\frac{1}{4}g_3\sigma^4-g_s\overline{\psi}_i\psi_i\sigma \nonumber\\
&&-{\frac{1}{4}}\Omega^{\mu\nu}\Omega_{\mu\nu}+{1\over{2}}m_{w}^{2}V^{\mu}V_{\mu} \nonumber\\
&&-g_\omega\overline{\psi}_i\gamma^{\mu}\psi_iV_{\mu}-{\frac{1}{4}}\vec{B}^{\mu\nu}.\vec{B}_{\mu\nu} \nonumber\\
&&+{1\over{2}}m_{\rho}^{2}{\vec R^{\mu}}.{\vec{R}_{\mu}}-g_\rho\overline{\psi}_i\gamma^{\mu}\overrightarrow{\tau}\psi_i.\overrightarrow{R^{\mu}} \nonumber\\
&&-{1\over{4}}F^{\mu\nu}F_{\mu\nu}-e\overline{\psi}_i\gamma^{\mu}\frac{(1-\tau_{3i})}{2}\psi_iA_{\mu}.
\end{eqnarray}
Here $\sigma$, $V_{\mu}$ and $\overrightarrow{R}_{\mu}$ are the fields for $\sigma$-,
 $\omega$- and $\rho$-meson respectively. $A^{\mu}$ is the electromagnetic
field. The $\psi_i$ are the Dirac spinors for the nucleons whose third
component of isospin is denoted by $\tau_{3i}$. $g_s$, $g_\omega$,
$g_\rho$ and $\frac{e^2}{4\pi}=\frac{1}{137}$ are the coupling constants
for the linear term of $\sigma$-, $\omega$- and $\rho$-mesons and photons
respectively. $g_2$ and $g_3$ are the parameters for the non-linear
terms of the $\sigma$-meson.
M, $m_\sigma$, $m_\omega$ and $m_\rho$ are the masses of the nucleons,
$\sigma$-, $\omega$- and $\rho$-mesons, respectively. $\omega^{\mu\nu}$,
$\overrightarrow{B}^{\mu\nu}$ and $F^{\mu\nu}$ are the field tensors for
the $V^{\mu}$, $\overrightarrow{R}^{\mu}$ and photon fields, respectively.
The quadrupole moment deformation parameter $\beta_2$, root mean square radii
and binding energy are evaluated using the standard relations 
\cite{pannert1987}. The nuclear density $\rho={\sum}^{A}_{i=1}\psi^{\dag}_i\psi_i$ is obtained by
solving the equation of motion obtained from the above Lagrangian
The values of the parameters for NL3 are given as \cite{lala97}
$g_{s}$=10.217, $g_{\omega}$=12.868,
$g_{\rho}$=4.574, $g_{2}$ = -10.431 ($fm^{-1}$), $g_3$ = -28.885, and M=939,
$m_{\sigma}$=508.194, $m_{\omega}$=782.501, $m_{\rho}$=763.0 in MeV.\\
\subsection{Total nuclear reaction cross section}
The details to calculate $\sigma_R$ using Glauber approach has been given
by R. J. Glauber \cite{gla59}. This model is based on the independent,
individual nucleon-nucleon ($NN$) collisions along the Eikonal approximation \cite{abu03}.
It has been used extensively
to explain the observed total nuclear reaction
cross-sections for various systems at high energies. The standard Glauber form
for the total reaction cross-sections at
high energies is expressed as \cite{gla59,kar75}:
\begin{equation}
\sigma_R=2\pi\int_0^\infty \textbf{b}[1-T(\textbf{b})]d\textbf{b},
\end{equation}
where `$T(\textbf{b})$' is the transparency function at impact parameter
`\textbf{b}'. The transparency function $T(\textbf{b})$ can be expressed in terms of phase shift
function as
\begin{equation}
T(\textbf{b})=|e^{\iota \chi_{PT}(\textbf{b})}|^2.
\end{equation}
Where $\chi_{PT}$ is the projectile-target phase shift function. This phase shift function
is calculated by
\begin{equation}
\iota\chi_{PT}(\textbf{b})=-\sum_{i,j}\sigma_{NN}\int\overline{\rho_{P}}(\textbf{s})
\overline{\rho_{T}}(|\textbf{b}-\textbf{s}|) d\textbf{s}.
\end{equation}
Here, the summation run over nucleons $i$ and $j$, where $i$ belongs to projectile
and $j$ belongs to target nuclei.
The subscript `$P$' and `$T$' refers to projectile and target
respectively. $\sigma_{NN}$ is the experimental nucleon-nucleon
reaction cross-section which depends on the energy. The z-
integrated densities are defined as
\begin{equation}
\overline{\rho}(w) =\int_{-\infty}
^\infty\rho(\sqrt{w^{2}+z^{2}}) d\textbf{z},
\end{equation}
with $w^2=x^2+y^2$.
The original Glauber model was designed for the
high energy approximation. However, it was found to work reasonably
well for both the nucleus-nucleus reaction and the differential
elastic cross-sections over a broad energy range \cite{chauvin1983,buenerd1984} by modified phase shift function as
\begin{equation}
\iota\chi_{PT}(\textbf{b})=-\int_P\int_T\sum_{i,j}[\Gamma_{NN}({b}_{eff})
 \overline{\rho_{P}}(\textbf{t})
\overline{\rho_{T}}(\textbf{s})]
d\textbf{s}d\textbf{t}.\\
\end{equation}
Where ${b}_{eff}$=$|\textbf{b}-\textbf{s}+\textbf{t}|$, $\textbf{b}$ is the impact parameter, in which
$\textbf{s}$ and $\textbf{t}$ are the dummy variables for integration
over the z-integrated target and projectile densities.\\
The phase shift function for projectile-target nucleus required the
multidimensional integrations. The Monte Carlo technique has been used for the evaluation of
realistic wave function which contain nuclear correlation. The optical limit approximation (OLA)
is used for the integration involving coordinates of projectile and target nuclei.\\
The phase shift function can be expressed for Glauber model many body (core+neutron) systems as
\begin{equation}
\iota\chi_{PT}=\iota\chi_{CT}+\iota\chi_{NT}.
\end{equation}
The core-target phase shift function ($\iota\chi_{CT}$) and nucleon-target phase shift
function ($\iota\chi_{NT}$) are defined in terms of densities of core(nucleon) and target by
\begin{equation}
\iota\chi_{CT}(\textbf{b})=-\int_C\int_T[\Gamma_{NN}(b_{eff})
 \overline{\rho_{C}}(\textbf{t})
\overline{\rho_{T}}(\textbf{s})]
d\textbf{s}d\textbf{t}\\
\end{equation}
and
\begin{equation}
\iota\chi_{NT}(\textbf{b})=-\int_T \Gamma_{NN}(|\textbf{b}-\textbf{s}|)\overline{\rho_{T}}(\textbf{s})
d\textbf{s}.
\end{equation}
The projectile-target phase shift function for many body system has been estimated by relation
\begin{equation}
e^{\iota\chi_{PT}}\rightarrow\langle\varphi_0|e^{\iota\chi_{CT}(\textbf{b})}+e^{\iota\chi_{NT}(\textbf{b})}|\varphi_0\rangle
\end{equation}
Here the $\varphi_0$ is the valence nucleon wave function.
Where the projectile-target profile function $\Gamma_{NN}$ for optical limit approximation is
defined as
\begin{equation}
\Gamma_{NN}({b}_{eff}) =\frac{1-\iota\alpha_{NN}}{2\pi
\beta^2_{NN}}\sigma_{NN} exp(-\frac{{b}^2_{eff}}{2\beta^2_{NN}}),
\end{equation}
for finite range and
\begin{equation}
\Gamma_{NN}({b}_{eff}) =\frac{1-\iota\alpha_{NN}}{2}\sigma_{NN} \delta ({b}_{eff}),
\end{equation}
for zero range.\\
The parameters $\sigma_{NN},\alpha_{NN}$ and $ \beta_{NN}$ usually
depend upon the proton-proton, neutron-neutron and proton-neutron
interactions. Here $\sigma_{NN}$ is the total nuclear reaction cross section of NN
collision, $\alpha_{NN}$ is the ratio of the real to the imaginary part
of the forward nucleon-nucleon scattering amplitude and $\beta_{NN}$ is the slope
parameter. The slope parameter determines the fall of the angular
distribution of the N-N elastic scattering.\\
It is worth mentioning that the
result in Glauber model is sensitive to the in-medium NN cross-section with
proper treatment of the input densities \cite{hussein91} and also depends on the
accuracy of the profile function.
At intermediate energies, medium effects can be taken into account on nucleon-
nucleon cross-sections. In NN scattering the basic input is the NN elastic
t-matrix. This t-matrix is modified to take into account nuclear medium effects in both
projectile and target. A. Bertulani et. al \cite{bertulani10} have shown that the nucleon knockout
reactions involving halo nuclei are more sensitive to medium modifications compared to normal nuclei.
Hence in our calculation we have used the Pauli blocking to get densities of odd nuclei.
The deformed or spherical nuclear densities obtained from the RMF model are
fitted to a sum of two Gaussian functions with
suitable co-efficient $c_i$ and ranges $a_i$ chosen for the respective nuclei which is expressed as
\begin{equation}
\rho (r)=\sum\limits_{i=1}^{2}c_{i}exp[-a_{i}r^{2}].
\end{equation}
The deformed intrinsic RMF densities are converted to its spherical equivalent
using this equation which is consistent with the Glauber theory applied in the
laboratory frame \cite{abu03}.
Then, the Glauber model is used to calculate the total reaction cross-section for both the stable
and unstable nuclei considered in the present study.  In Refs: \cite{panda2009,abu03,pshukla03,bhagwat} 
clearly seen that the Glauber model can
be used for relatively low energy even at 25, 30 and 85 MeV/nucleons. Although
it is a better prescription to take deformation into account directly through
the transparency function [Eqn. (3)], but in our knowledge no such scheme is
available in this model. In our earlier calculations
\cite{skpatra1,panda2009,sharma06,mahesh2016} we have used the present approach
to take deformation into account where the
results are quite encouraging and show clear deformation effect. It is to be
noted that similar methodology is also adopted by some other authors
\cite{bhagwat}. Also, it is important to note that the densities for
halo-nuclei have long tails which generally are not reproduced quantitatively by harmonic
oscillator expansion in a mean field formalism.\\
\subsection{Angular elastic differential cross section}
The nucleus-nucleus elastic scattering amplitude is
written as
\begin{equation}
F(\textbf{q})=\frac{\iota K}{2\pi} \int d\textbf{b} e^{\iota \textbf{q}.\textbf{b}}(1-e^{\iota \chi_{PT}(\textbf{b})}).
\end{equation}
At low energy,  this model is modified in order to take care of finite range
effects in the profile function and Coulomb modified trajectories. The elastic
scattering amplitude including the Coulomb interaction is expressed as
\begin{equation}
F(\textbf{q})=e^{\iota \chi_{s}}\{F_{coul}(\textbf{q})+\frac{\iota K}{2\pi}
\int d\textbf{b} e^{\iota \textbf{q}.\textbf{b}+2\iota \eta \ln(Kb)}(1-e^{\iota \chi_{PT}(\textbf{b})})\},
\end{equation}
with the Coulomb elastic scattering amplitude
\begin{equation}
F_{coul}(\textbf{q})=\frac{-2 \eta K}{q^2}exp\{-2 \iota \eta \ln(\frac{q}{2K})
+2\iota arg \Gamma(1+\iota \eta)\},
\end{equation}
where $K$ is the momentum of projectile and $q$ is the momentum transferred from the projectile to the target.
Here $\eta=Z_P Z_T e^2/\hbar v$ is the
Sommerfeld parameter, $v$ is the incident velocity of the projectile, and $\chi_s=-2\eta \ln(2 K a)$ with $a$
being a screening radius.
The elastic differential cross section is given by
\begin{equation}
\frac{d\sigma}{d\Omega}=|F(\textbf{q})|^2.\\
\end{equation}
Whereas the ratio of angular elastic to the Rutherford elastic differential cross section
is expressed as
\begin{eqnarray}
\frac{d\sigma}{d\sigma_R}=\frac{\frac{d\sigma}{d\Omega}}{\frac{d\sigma_R}{d\Omega}}=\frac{|F(\textbf{q})|^2}{|F_{coul}(\textbf{q})|^2}.
\end{eqnarray}
\subsection{One nucleon removal cross section}
The expression for one nucleon removal reaction cross-section $\sigma_{-N}$
is given by \cite{abu03}
\begin{equation}
\sigma_{-N}=\sum_{c}\int d\overrightarrow{k}\sigma_{a=(k,g=0),c},
\end{equation}
\noindent where $\sigma_{a=(k,g=0),c}$ are the possible final states $ac$.
In the present formalism, it is considered that the projectile nucleus breaks
up into a core and the removed nucleon. The core $c$ has an internal wave
function $\phi_g$ and the one-nucleon, \textit{i.e}., the $"picked-up"$ nucleon has an
asymptotic momentum $\hbar {\bf k}$ in the continuum state with respect to
the core.  The core is considered to be in the ground state ($g=0$) at the time of the collision.
The total $\sigma_{-N}$ can be separated to an elastic
($\sigma_{-1N}^{el}$) with $c=0$ and inelastic ($\sigma_{-1N}^{iel}$) part
having $c$ as non-zero. The $\sigma_{-1N}^{el}$ and $\sigma_{-1N}^{iel}$ are
expressed as \cite{abu03}
\begin{eqnarray}
\sigma_{-1N}^{el}&=&\int d{\bf b}\{<\phi_0\mid e^{-2Im\chi_{Ct}(\bf b_C)
-2Im\chi_{nt}(\bf b_C+\bf s)}\mid \phi_0> \nonumber \\
&&-\mid <\phi_0\mid e^{-i\chi_{Ct}(\bf b_C)
+i\chi_{nt}(\bf b_C+\bf s)}\mid \phi_0>\mid^2\},
\end{eqnarray}
\begin{eqnarray}
\sigma_{-1N}^{iel}&=&\int d{\bf b}\{<\phi_0\mid e^{-2Im\chi_{Ct}(\bf b_C)}
\nonumber \\
&&-e^{-2Im\chi_{Ct}(\bf b_C)-2Im\chi_{nt}(\bf b_C+\bf s)}
\mid \phi_0>\}.
\end{eqnarray}
\noindent Here $\phi_0$ is the valence wave function (the wave function of the
removed nucleon),  $\bf b_C$ is the impact parameter between the core and the
target, $\chi_{Ct}$, $\chi_{nt}$ are the core-target and nucleon-target phase-
shift functions respectively and $\bf s$ is the coordinate of the nucleon with
respect to the projectile/target.
The notation and the numerical procedure of calculation for one-nucleon removal
reaction cross-section are from Ref. \cite{abu03}.
It is worth mentioning that the Eikonal theory is based on centre of mass system whereas we
have calculated  $\sigma_{-N}$ in the laboratory frame. From the
analysis of stripping and diffractive single-neutron removal cross-section, one can find
\cite{batham05} that the diffractive cross-section increases with
deformation  $\beta_{2}$. The cross-section of the stripping reaction remains
almost insensitive with $\beta_{2}$. In our present study we use the density of the actual
deformation calculated by NL3 parameter set. Eikonal approximation is useful to investigate
the effects of the dynamical reorientation in the
single nucleon knockout reaction. Interestingly, we investigate the deformation
effect of the participating nuclei (both target and projectile) using Glauber
model. This includes the actual structure obtained through the RMF deformed
densities.\\
\subsection{Longitudinal momentum distribution}
The momentum distribution of core fragment after the inelastic breakup of projectile can be written as:
\begin{eqnarray}
\frac{d\sigma^{inel}_{-N}}{d\textbf{P}}=\int \frac{d\textbf{q}}{K^2} \sum_{c\neq0}\int d\textbf{k} \delta (\textbf{P}-\frac{A_C}{A_P} \hbar \textbf{q}+\hbar \textbf{k} )|F_{(k,0)c} (\textbf{q})|^2 .
\end{eqnarray}
Where the momentum of the core is expressed as $\textbf{P}=(P_{\parallel}, \textbf{P}_\perp$) and the nucleon going to continuum state be $\hbar\textbf{k}$. Scattering amplitude for the reaction in continuum state is $F_{(k,0)c}\textbf{(q)}$. The core-nucleon scattering wave function is approximated by a plane wave and we assume that the core remains in its ground state. This reduces the above expression to the form
\begin{eqnarray}
\frac{d\sigma^{inel}_{-N}}{d\textbf{P}}=\int d\textbf{b}_N (1-e^{-2Im \chi_{NT}(\textbf{b}_N)})\times\frac{1}{(2\pi\hbar)^3} \frac{1}{2j+1} \nonumber\\
\sum_{m m_s}|\int d\textbf{r} e^{\frac{i}{h}\textbf{P}.\textbf{r}}\chi_{\frac{1}{2}m_s} e^{i\chi_{CT}(\textbf{b}_N-\textbf{s})}\varphi_{nljm}(\textbf{r})|^2,
\end{eqnarray}
where $\textbf{b}_N$ stands for the impact parameter of valence nucleon with respect to the target and $\varphi_{nljm}(\textbf{r})$ is
the wave function of valence nucleon which may expressed as
\begin{eqnarray}
\varphi_{nljm}(\textbf{r})=u_{nlj}(r)\sum_{m_l m_s}\langle l m_l\frac{1}{2} m_s|jm\rangle Y_{l m_l}(\hat{\textbf{r}}) \chi_{\frac{1}{2} m_s}.
\end{eqnarray}
The longitudinal momentum distribution obtained by the integration of above equation (23) over transverse component of
momentum ($\textbf{P}_{\bot}$), gives.
\begin{eqnarray}
\frac{d\sigma^{inel}_{-N}}{dP_{||}}=\int d\textbf{P}_{\bot}\frac{d\sigma^{inel}_{-N}}{d\textbf{P}}\nonumber
\end{eqnarray}
\begin{eqnarray}
&=&\frac{1}{2\pi\hbar}\int d\textbf{b}_N(1-e^{-2I_m \chi_{NT}(\textbf{b}_N)})\int d\textbf{s} e^{-2I_m \chi_{CT}(\textbf{b}_{N}-\textbf{s})}\nonumber\\
&& \times \int dz \int dz' e^{\frac{\iota}{\hbar}P_{||}(z-z')} u^{*}_{nlj}(r') u_{nlj}(r) \frac{1}{4\pi} P_l(\hat{\textbf{r}}'.\hat{\textbf{r}}).
\end{eqnarray}
here $\textbf{r}=(\textbf{s},z)$ and $\textbf{r}'=(\textbf{s},z')$ and $P_l$ is the Legendre Polynomial.\\
\section{Results and Discussions}
We obtain the field equations for nucleons and mesons from the RMF Lagrangian.
For deformed case, these equations have been solved by
expanding the upper and lower components of the Dirac spinners and the boson
fields in an axially deformed harmonic oscillator basis. A set of coupled
equations have been solved numerically by a self-consistent iteration method taking
different inputs of the initial deformation $\beta_0$
\cite{patra91,serot86,pring96,gam90}. For spherical densities, we have followed the
same numerical
procedure as of Refs: \cite{patra01,skpatra01} for RMF model.
The center-of-mass motion (c.m.) energy
correction is estimated by the usual harmonic oscillator formula
$E_{c.m.}=\frac{3}{4}(41A^{-1/3})$.\\
\subsection {Density, binding energy, charge radius and quadrupole deformation}
The ground state properties, i.e. binding energy (BE), the rms charge radius ($r_{ch}$) and quadrupole deformation parameter ($\beta_2$) for Al isotopes starting from proton-rich $^{22}$Al to the expected drip-line nuclei $^{44}$Al and
$^{12}$C are being estimated within the RMF approximation. The
calculated results are presented in Table I, obtained from both spherical and
axially deformed RMF formalisms. The experimental data \cite{CPC12} are also
given for comparison. There is an overall good agreement between our results and experimental data for all the isotopic chains of Al. A careful analysis show that, upto $^{30}$Al, the calculated value under estimates the experimental data, whereas it over estimates the data henceforth. Comparing the binding energies (BE) of the calculated
solutions, the maximum BE  and the corresponding densities
[$\rho_p$ (proton) and $\rho_n$ (neutron)] are the ground state properties, whereas rest solutions are the excited intrinsic state even including the spherical one.
Since the main inputs of the Glauber model estimation are the densities of projectile and target nuclei, it is important to have information about these quantities. We have
plotted both spherical proton ($\rho_p$) and neutron ( $\rho_n$) density distributions of
$^{23-28}$Al isotopes in Fig. 1. Here the nucleonic density distribution is of
 maximum amplitude at $\sim$ 2 fm, whereas it decreases slightly toward the center and then go on decreasing with increase in radius from 2 fm. A dip-like density structure is appeared at the centre of the nucleus depicting bubble like structure of these nuclei. As expected, we
find an extended density distribution for proton as compared to
neutron for $^{23-25}$Al, which is due to the proton-rich nature of these nuclei.
The values of $\rho_n$ and $\rho_p$ are almost similar for $^{26}$Al as seen in figure 1.
Extension of $\rho_n$ is much more than $\rho_p$ for rest considered nuclei.  It is
maximum for $^{28}$Al in isotopic chains, because of high neutron
to proton ratio for this case. One may also observe from the figure that the
skin effect increases with increase in isotopic mass number.\\

\begin{figure}
\hspace{0.9cm}
\includegraphics[width=1.0\columnwidth,clip=true]{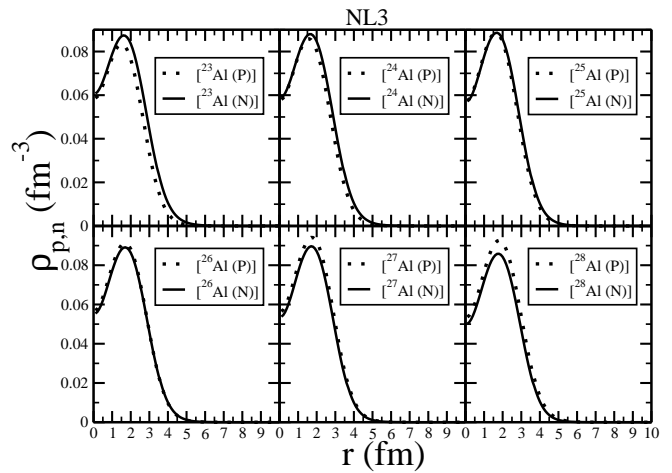}
\label{Fig.1}\vspace{-0.7cm} \caption{The spherical proton ($\rho_p$) and neutron
($\rho_n$) density obtained from RMF (NL3) parameter set for various
isotopes of Al}
\end{figure}
The observed rms charge radius $r_{ch}$ of all the isotopes of Al for both spherical and deformed
cases in RMF(NL3) are also presented in Table 1. The experimental values \cite{angeli2013} are
given for comparison. The rms charge radius ($r_{ch}$) is obtained through the relation
$r_{ch} = \sqrt{{r_p}^2 + 0.64}$
\cite{patra91,gam90},The observed rms charge radius $r_{ch}$ of all the isotopes of Al for both spherical and deformed
cases in RMF(NL3) are also presented in Table 1. The experimental values \cite{angeli2013} are
given for comparison. The rms charge radius ($r_{ch}$) is obtained through the relation
$r_{ch} = \sqrt{{r_p}^2 + 0.64}$
\cite{patra91,gam90}, where the factor 0.64 indicates the finite size effects of protons with radius 0.8 fm. It can be noticed that, the result obtained here slightly overestimates the experimental data.
\begin{table*}
\caption{\label{tab:table3}Binding energy (BE), charge radius ($r_{ch}$) and
quadrupole deformation parameter ($\beta_2$) for the projectile $^{22-44}$Al
isotopes and the target $^{12}$C obtained from (RMF)NL3 parameter set are compared with the experimental results whichever is available. The BE is in MeV and charge radius in fm.}
\begin{tabular}{|c|c|c|c|c|c|c|c|c|c|c|c|c|c|c|c|c|c|}
\hline \hline
Nucleus &\multicolumn{3}{c|}{BE}&\multicolumn{3}{c|}{$r_{ch}$}&\multicolumn{2}{c|}
{$\beta_2$}\\
\hline
&NL3(Sph.)&NL3(Def.)& Expt.\cite{CPC12} &NL3(Sph.)&NL3(Def.)&
Expt.\cite{angeli2013}&NL3(Def.)& Expt.\cite{nndc}\\ \hline
$^{22}$Al&152.345&150.222&149.204&3.22&3.138&&0.284& \\
&&149.371&&&3.128&&-0.200&\\ \hline
$^{23}$Al&166.430&167.382&168.721&3.132&3.128&&0.399&0.308 \\
&&165.154&&&3.103&&-0.261&\\ \hline
$^{24}$Al&180.917&182.219&183.589&3.126&3.098&&0.388&\\
&&179.295&&&3.078&&-0.258&\\ \hline
$^{25}$Al&194.745&197.623&200.529&3.062&3.073&&0.382& \\
&&193.799&&&3.061&&-0.266&\\ \hline
$^{26}$Al&208.077&206.522&211.894&3.192&3.053&&0.5498&0.313 \\
&&207.694&&&3.053&&-0.275&\\ \hline
$^{27}$Al&221.190&216.904&224.91&3.001&3.216&3.06&0.572&0.448 \\
&&221.729&&&3.054&&-0.293&\\ \hline
$^{28}$Al&230.376&229.793&232.677&3.020&3.031&&0.206&0.254 \\
&&229.386&&&3.038&&-0.208&\\ \hline
$^{30}$Al&247.447&246.007&247.84&3.069&3.07&&0.207&0.207 \\
&&245.459&&&3.071&&-0.184&\\ \hline
$^{32}$Al&261.683&259.844&259.210&3.103&3.10&&0.113&0.000 \\
&&259.629&&&3.103&&-0.111&\\ \hline
$^{34}$Al&272.243&269.724&267.323&3.131&3.139&&0.59&0.000 \\
&&269.268&&&3.134&&-0.107&\\ \hline
$^{36}$Al&278.598&277.413&274.446&3.150&3.187&&0.313&0.234 \\
&&274.904&&&3.174&&-0.189&\\ \hline
$^{38}$Al&283.521&283.042&280.33&3.170&3.216&&0.3388&0.273 \\
&&280.172&&&3.214&&-0.261&\\ \hline
$^{40}$Al&289.546&287.410&284.72&3.189&3.305&&0.44&0.189 \\
&&286.542&&&3.258&&-0.3357&\\ \hline
$^{42}$Al&292.727&290.497&287.994&3.208&3.348&&0.482&0.26 \\
&&290.425&&&3.281&&-0.354&\\ \hline
$^{44}$Al&295.160&290.775&&3.227&3.375&&0.454&0.23 \\
&&291.991&&&3.274&&-0.287&\\ \hline
$^{12}$C&91.19&91.010&92.2&2.36&2.693&2.47&0.118&0.577 \\
&&90.782&&&2.68&&-0.136\\
&&90.672&&&2.682&&0.0007&\\ \hline
\end{tabular}
\end{table*}

The quadrupole deformation parameter ($\beta_2$) is determined to know the shape of the nuclei. It is being calculated from the RMF formalism and compared with the available data \cite{nndc} as shown in Table I . It is found that the results are in line with the data. Moreover for $^{30}$Al, the calculated value of $\beta_2$ is exactly equal to that of the experimental data, i.e. $\beta_2$ = 0.207. The positive $\beta_2$ values are of prolate and
negative for oblate deformation. It is to be noted that present parameter set NL3 does not provide a converged solution for many light mass nuclei. We slightly change the strength $\triangle_{n,p}$ of BCS pairing in order to get a converged result. Here, we compromise a small bit in quadrupole deformation calculation.
The quadrupole deformation parameter ($\beta_2$) is determined to know the shape of the nuclei. It is being calculated from the RMF formalism and compared with the available data \cite{nndc} as shown in Table I . It is found that the results are in line with the data. Moreover for $^{30}$Al, the calculated value of $\beta_2$ is exactly equal to that of the experimental data, i.e. $\beta_2$ = 0.207. The positive $\beta_2$ values are of prolate and
negative for oblate deformation. It is to be noted that present parameter set NL3 does not provide a converged solution for many light mass nuclei. We slightly change the strength $\triangle_{n,p}$ of BCS pairing in order to get a converged result. Here, we compromise a small bit in quadrupole deformation calculation.

\subsection {Proton, neutron separation energy and wave function}
To understand the definite structure and their effects on the nuclei, we have studied the one and two proton separation energy ($S_p,S_{2p}$) and one-neutron
separation energy ($S_{n}$) with respect to mass number. These are being shown in Fig. 2. In this figure we have compared our results with the well known finite range droplet model (FRDM) \cite{FRDM}, and with the available experimental data \cite{CPC12}.

Ref. \cite{wapstra1985} suggests a proton separation energy of
5.497 MeV for $^{22}$Mg, where as $S_p$ is of 4.917 MeV from our calculations.
So $^{22}$Mg is possibly a good inert core in $^{23}$Al, which is evident from the large proton separation energy. Thus our result supports Ref. \cite{wapstra1985} that there can be a proton halo in $^{23}$Al. The proton separation energy ($S_p$) of $^{24}$Al is found to be 2.283 MeV where as 1.871 MeV from
experimental data. This result satisfies the assumption of proton skin.
It can be seen that with the increase of mass number, $S_p$ increases gradually. This shows that $^{23}$Al is weakly bound as compared to other in the isotopic
chain. We have observed the similar  behaviour in $S_{2p}$ as that of the $S_p$, with the increase of mass number toward the drip-line its value increases
gradually. One-neutron separation energy ($S_{n}$) decreases in order with
increase in mass number.\\

\begin{figure}
\vspace{0.5cm}
\includegraphics[width=1.0\columnwidth,clip=true]{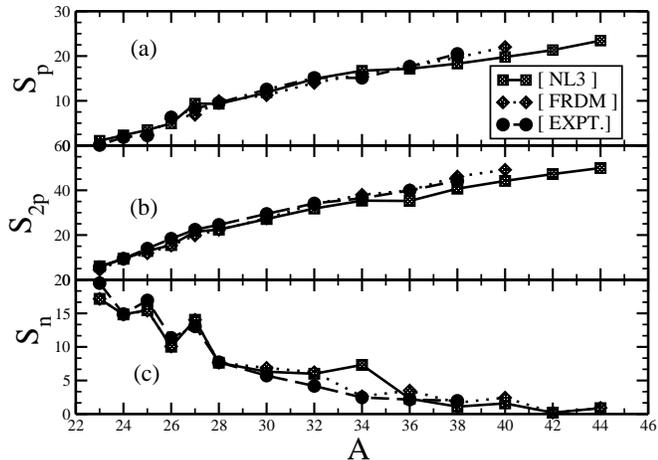}
\label{Fig.1}\vspace{-0.7cm} \caption{Represent one
proton, two proton and one neutron separation energy for various isotopes of
aluminium obtained from NL3 parameter set for deformed densities compared with
FRDM model and experimental data.}
\end{figure}
\begin{table}
\caption{\label{tab:table3}Different observables for $^{23}$Al obtained by RMF calculation using
NL3 parameter set with available experimental data.}
\begin{tabular}{|c|c|c|c|c|c|c|c|c|}
\hline \hline
Observables&RMF(NL3)&Ref. \cite{gen2004}& Exp.\\ \hline
B.E. (MeV)&167.382&166.27&168.721\\
$\beta_p$&0.391&0.53&\\
$\beta_n$&0.412&0.43&\\
$\beta_2$&0.399&0.52&0.308\\
$r_m$(fm) &2.925&3.14&2.91$\pm 0.25$\\
$r_p$(fm) &3.026&3.26&3.1$\pm 0.25$\\
$r_n$(fm) &2.789&2.97&2.634$\pm 0.23$\\
$S_p$(MeV) &1.08&1.90&0.122\\
$S_{2p}$(MeV)&5.993&5.13&5.645\\
\hline
\end{tabular}
\end{table}
\begin{figure}
\vspace{0.5cm}
\includegraphics[width=1.0\columnwidth,clip=true]{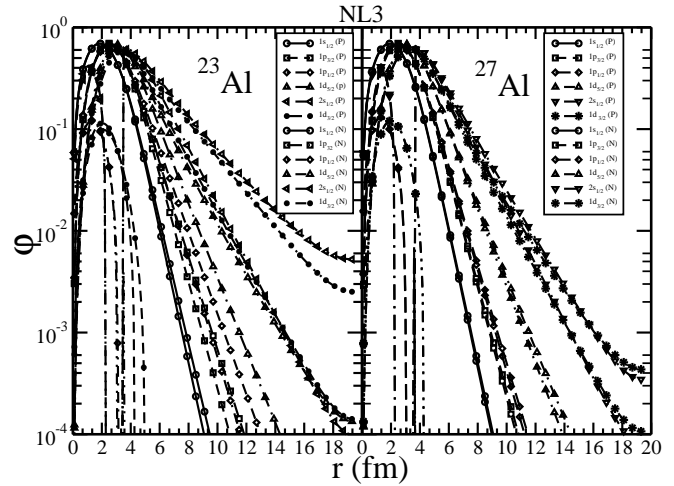}
\label{Fig.1}\vspace{-0.7cm} \caption{Comparison of wave function
distribution at different energy levels for both proton and neutron in
$^{23}$Al and $^{27}$Al.}
\end{figure}
\begin{table*}
\caption{\label{tab:table3}The Gaussian coefficients $c_i$'s and ranges $a_i$'s for the projectiles and targets using RMF (NL3) spherical and deformed densities.}
\begin{tabular}{|c|c|c|c|c|c|c|c|c|}
\hline \hline
Nucl. &\multicolumn{4}{c|}{RMF(NL3) Sph.}&\multicolumn{4}{c|}{RMF(NL3) Def.}\\
\hline
$^{23}$Al&-3.81017&0.288261&3.92516&0.265158&-2.41626&0.264886&2.66882&0.24618\\ \hline
$^{24}$Al&-4.09865&0.283709&4.20988&0.261782&-4.21864&0.310612&4.52894&0.292168\\ \hline
$^{25}$Al&-4.17663&0.280378&4.28444&0.258387&-6.36213&0.364051&6.74406&0.343821\\ \hline
$^{26}$Al&-4.3287&0.277001&4.43338&0.255346&-4.24438&0.398225&4.56748&0.355406\\ \hline
$^{27}$Al&-4.47652&0.274111&4.57837&0.252709&-4.07396&0.369559&4.4114&0.332769\\ \hline
$^{28}$Al&-4.41178&0.262791&4.50695&0.242354&-3.90986&0.366962&4.2608&0.328647\\ \hline
$^{12}$C&-0.232654&0.6388687&0.517232&0.339911&-1.14333&0.285974&1.47032&0.285598 \\ \hline
\end{tabular}
\end{table*}

While studying different observable as given in Table II, we found a better fitting of our results with experimental data as compared to results of reference \cite{gen2004}. The calculated binding energy (BE) for $^{23}$Al is 167.382 MeV, which is in agreement with the experimental data of 168.721 MeV \cite{CPC12} as well as with the result 166.27 MeV of Ref. \cite{gen2004}. The relative error between our result and the data is about 0.8\%. Similarly, the quadrupole deformations for proton $\beta_p$ and neutron $\beta_n$ are 0.391 and 0.412, where as Ref. \cite{gen2004} results are 0.53 and 0.43 respectively. Hence the nucleus $^{23}$Al possesses indeed a very large
deformation. The observed rms radii of matter, proton and neutron are in excellent agreement with the data. The predicted one-proton and two-proton separation energies are somewhat overestimated by the RMF(NL3) as compared with the data \cite{audi1993a}.\\ 

To get further understanding about the formation of the halo, the comparison of the wave function of the particles for different energy levels of both proton and neutron are plotted in Fig. 3. This figure shows variation of wave function with radial distance. In both the cases ($^{23}$Al and $^{27}$Al), the inner orbitals are almost similar to each other, however, the two outermost levels (2s and $1d_{5/2}$) are very different than the $^{27}$Al. In case of $^{27}$Al all the levels are in compact form ruling out the formation of halo or skin type structure. On the other hand the $2s_{1/2}$ and $1d_{5/2}$ protons have an extended area of existence, giving the impression of the halo formation. Thus, we observed a long extension of the wave function for the proton rich $^{23}$Al as compared to the stable isotope $^{27}$Al, suggesting the possibility of a halo structure of $^{23}$Al nucleus.\\

\subsection {Total reaction, differential and one proton removal cross section}
In the measurement of reaction parameters through Glauber formalism, one of the input for evaluation of profile function is its energy as well as isospin dependent parameters. The values of these parameters at $E_{Proj.}$= 30 and 74 MeV/nucleon are $\sigma_{NN}$= 19.6, 6.893758 $(fm^2)$, $\alpha_{NN}$=0.87, 1.255045 and $\beta_{NN}$= 0.685, 0.3539421 $(fm^2)$ respectively, as estimated from Ref. \cite{horiu07}.\\

\begin{figure}
\hspace{0.9cm}
\includegraphics[width=1.0\columnwidth,clip=true]{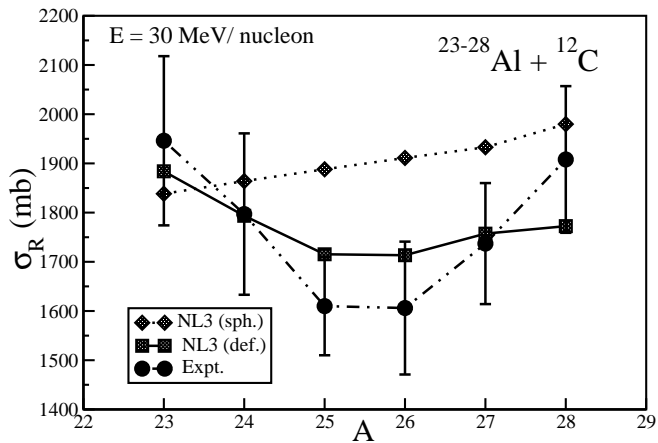}
\label{Fig.1}\vspace{-0.7cm} \caption{shows the variation of Coulomb-modified reaction cross-sections $\sigma_R$ in mb for $^{23-28}$Al + $^{12}$C reactions with the experimental data.}
\end{figure}
\begin{figure}
\hspace{0.9cm}
\includegraphics[width=1.0\columnwidth,clip=true]{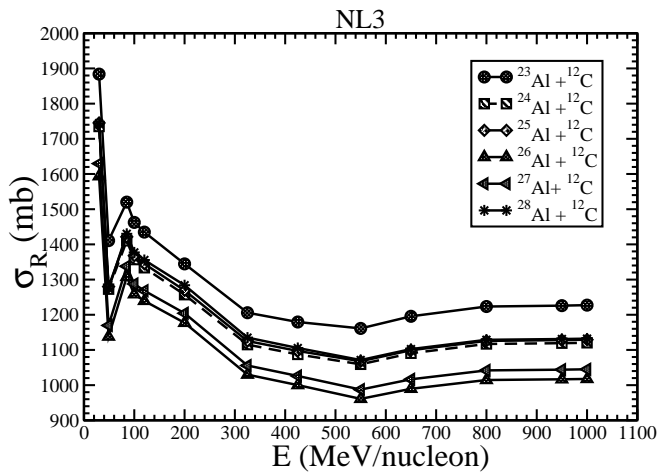}
\label{Fig.1}\vspace{-0.7cm} \caption{same as figure 4 but for projectile energy 30 - 1000 MeV/nucleon range.}
\end{figure}

\begin{figure}
\hspace{0.9cm}
\includegraphics[width=1.0\columnwidth,clip=true]{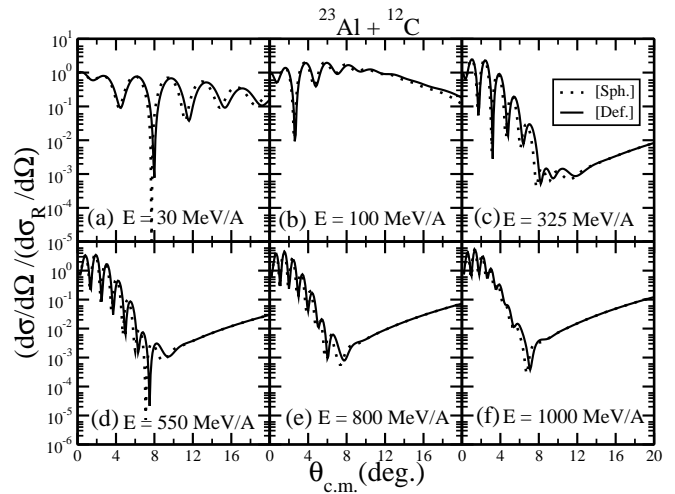}
\label{Fig.1}\vspace{-0.7cm} \caption{differential elastic scattering for
$^{23}$Al + $^{12}$C reactions at various projectile energy as a function of
scattering angle using RMF(NL3) formalism.}
\end{figure}

\begin{figure}
\hspace{0.9cm}
\includegraphics[width=1.0\columnwidth,clip=true]{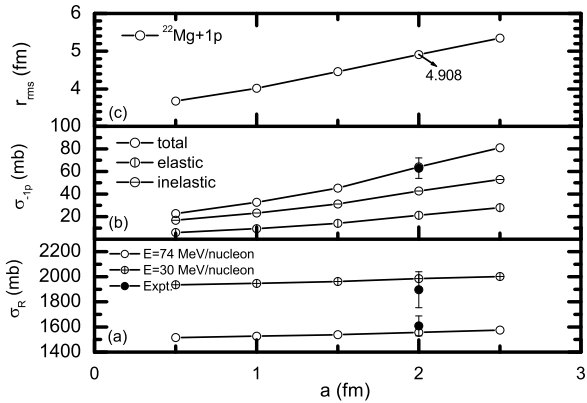}
\label{Fig.1}\vspace{-0.7cm} \caption{Upper panel of the figure show the rms
radius ($r_{rms}$) in fm (b) middle panel show one proton removal $\sigma_{-1p}$ in mb as a function diffuseness parameter `a' in fm for $^{23}$Al + $^{12}$C
reaction and (c) lower panel show the values of reaction cross section
$\sigma_R$ in mb as a function diffuseness parameter `a' in fm for
$^{23}$Al + $^{12}$C reaction.}
\end{figure}

 We have compared the reaction cross section $\sigma_R$ obtained from NL3 parameter set for both spherical and deformed densities of the projectiles $^{23-28}$Al and $^{12}$C target nuclei with the available experimental data \cite{cai2002}.
Fig. 4 shows the behaviour of reaction cross section at intermediate energy
30 MeV/nucleon for $^{23-28}$Al + $^{12}$C reactions with NL3 parameter, which are then compared with the available experimental data. If we go detail into the figure then it could be clearly noticed that the experimental values are more closer with the results obtained from  NL3 parameter set for deformed densities. In this figure we found that reaction cross sections increase with mass number A, for nuclei $^{26,27,28}$Al.
However, $\sigma_R$  increases  with decrease of A from
$^{25}$Al to $^{23}$Al as it approaches to the proton drip line. This trend indicates the appearance of proton skins or halos of these nuclei. Larger cross
section of $^{24}$Al as compared to $^{25}$Al indicates the possibility of proton skin in $^{24}$Al. Further, the enhanced cross section of $^{23}$Al as compared to its neighbouring nuclei is suggested either highly deformed
structure or proton halo nature of $^{23}$Al isotope. In Fig. 5, we have presented our calculated $\sigma_R$ values for $^{23-28}$Al
as projectiles and $^{12}$C as target for incident energy range 30-1000
MeV/nucleon. In this case, the total reaction cross section increases with
increase in mass of the projectile. However, at relatively low incident energies (30-200 MeV/nucleon), $\sigma_R$ is maximum and then it decreases with further
increase of energy. Thus the maximum value of $\sigma_R$ occurs at a particular
incident energy, irrespective of the mass of the projectile or target. In general, we observed that except $^{23-24}$Al+$^{12}$C reactions, the spherical RMF density fails to reproduce the experimental data. Whereas the deformed densities enhance the accuracy of  $\sigma_R$ values for the set of considered isotopes in comparison to experimental data.\\
Differential elastic scattering cross section are important to explain
scattering phenomenon. Fig. 6 shows our results for $^{23}$Al + $^{12}$C
system at energies 30, 100, 325, 550, 800 and 1000 MeV/nucleon. In this figure we found that the
differential elastic scattering cross section shows a large variation with
incident energy. At lower value of projectile energy diffraction dissociation is uniform, where as the dissociation go on decreasing with increase of projectile energy with increasing angles. Clear oscillatory pattern appeared at projectile energy 325 MeV/nucleon and above, amplitude of oscillations are disappeared at higher angle with increase of projectile energy.\\

In order to study the halo nature of exotic nuclei from cross section
measurement Ozawa et al. \cite{ozawa1996} suggested a difference factor $d$. If
$d$ in a nucleus is evidently larger than its neighbouring nuclei, there will appear nucleon halo in that nucleus. They successfully explain about the neutron halo in $^{15}C$ among carbon isotopes as well as in some other nuclei also. 
The difference factor $d$ is written as
\begin{equation}
d=\frac{\sigma_{R}(exp)-\sigma_{R}(cal)}{\sigma_{R}(cal)},
\end{equation}
where $\sigma_{R}(exp)$ and $\sigma_{R}(cal)$ are the experimentally and
theoretically calculated values of total nuclear cross section at intermediate
energy. According to our results the factor $d$ for proton rich $^{23}$Al is
0.12 where as for $^{24-28}$Al isotopes are 0.04, 0.077, 0.06, 0.01 and 0.09
respectively. Thus the factor $d$ for $^{23}$Al is order of magnitude larger than others. Therefore
we can conclude that $^{23}$Al is a proton halo nucleus.\\

\begin{figure}
\hspace{0.9cm}
\includegraphics[width=1.0\columnwidth,clip=true]{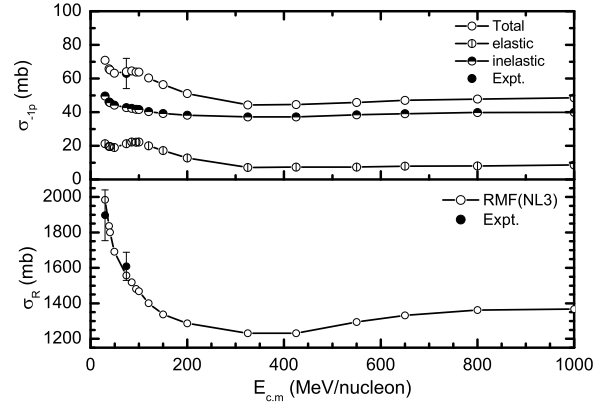}
\label{Fig.1}\vspace{-0.7cm} \caption{Upper panel show one proton removal
$\sigma_{-1p}$ in mb and lower panel reaction cross section $\sigma_R$ in mb
as a function diffuseness parameter `a' in fm for $^{23}$Al + $^{12}$C
reaction at projectile energy 0 to 1000 MeV/nucleon range with available
experimental data.}
\end{figure}

\begin{figure}
\vspace{0.3cm}
\hspace{0.5cm}
\includegraphics[width=1.0\columnwidth,clip=true]{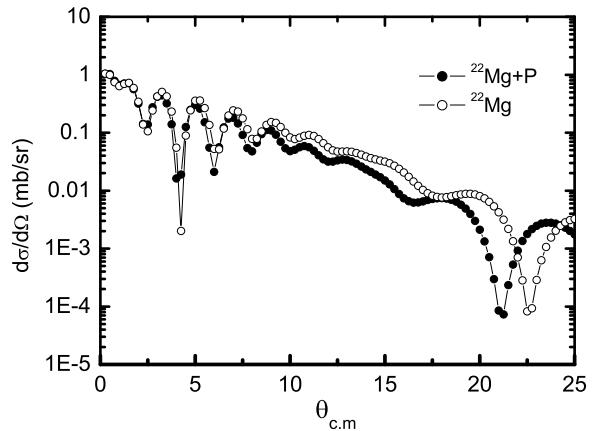}
\label{Fig.1}\vspace{-0.7cm} \caption{The comparison of angular elastic differential cross section
of $^{22}$Mg and $^{23}$Al as a function of scattering angle at projectile energy 74 MeV/nucleon.}
\end{figure}

As clear from Fig. 4, the estimated value of $\sigma_R$ from deformed RMF is 1884 mb, spherical RMF is 1838 mb and experimental value is 1946$\pm$172 mb. Hence the deformed RMF predicts closer results to the experimental data in comparison to the spherical density. Although the deformation plays significant role to enhance the accuracy in the results but still more analysis is needed for the description of higher experimental value of $\sigma_R$. Hence, in this concern variation of diffuseness effect have been investigated for more concluding analysis of $^{23}$Al.
For further inspection of $^{23}$Al, we have applied Glauber model
with many body (core+nucleon) system. The detailed calculations
are as from ref's \cite{abu03,mahesh2013,mahesh2016}. The Schrodinger equations have been solved using potential between core and valence nucleon 
for getting the solution in terms of radial part of single particle wave function.
The used potential is in the form of Wood Saxon as given below:
\begin{eqnarray}
U(r)=-v_{0}f(r)+V_{ls}(l.s)r_0^2\frac{1}{r}\frac{df(r)}{dr}+V_{Coul},
\end{eqnarray}
where $f(r)=[1+\exp(\frac{r-R}{a})]^{-1}$ and $R=r_0A^{(1/3)}$. The first term of equation (27) contains
the central potential, second term contains spin orbital part and the last term of the equation contains
Coulomb part of potential. $`A'$ be the mass number of nucleus.
We fixed the value of $r_0=1.7$ fm and
diffuseness parameter $`a'$ vary from $0.6-2.5$ fm in our calculations.\\

Fig. 7 represents the
calculated values of $\sigma_R$, $\sigma_{-1p}$ and $r_{rms}$ for the $^{23}$Al projectile. The upper panel of this figure represents the root mean square radius $r_{rms}$ for $^{23}$Al of core + nucleon system, where as middle panel
show $\sigma_{-1p}$ and lower panel $\sigma_{R}$ for $^{23}$Al+$^{12}$C reactions at energy 30 and 74 MeV/nucleon respectively as a function of diffuseness parameter `a' in fm. We have taken care to fit the reaction
cross section for $^{23}$Al at different values of diffuseness parameter. It is observed from the figure that both reaction cross section and diffuseness parameter are linearly dependent to each other. The values of $\sigma_R$ obtained at diffuseness parameter = 2 fm are 1950 and 1537 mb, for incident energy 30 and
74 MeV/nucleon respectively, where as the experimental values 1946$\pm$172 and
1609$\pm$79 mb. Therefore, the calculated value of reaction cross section fit
with the experimental data at a = 2 fm as shown in figure 7. Unlike the total reaction cross-section, the $\sigma_{-1p}$ obtained from the Glauber model depends very much on the structure information of the projectile and target nuclei, i.e. input densities of these systems. 

The proton removal reaction
cross-section of $^{23}$Al is 64 mb at 74 MeV/nucleon, which is well comparable with the predictions of Frag et al. \cite{frag2007} i.e the data of 63$\pm 9$mb,  larger than its neighbours, suggesting a weak binding of the last proton and extended valence density distribution. We also found the value of root mean square radius of
$^{23}$Al (core + nucleon) at a = 2 fm is 4.908 fm. Thus the large value of reaction cross section due to extended matter density distribution for $^{23}$Al isotope also supported our previous observations of halo status of such nuclei.\\

Fig. 8 represents the variation of $\sigma_R$ and $\sigma_{-1p}$ in mb as a function of projectile energy over the energy range 30-1000 MeV/nucleon for diffuseness 2 fm. The upper panel of the figure consists total, elastic and inelastic components of the $\sigma_{-1p}$ cross section with available data.
The elastic differential cross sections as a function of angular distribution
for both $^{22}$Mg core and $^{23}$Al halo projectiles on the carbon target at
$E_{Proj.}$= 74 MeV/nucleon is shown in Fig. 9. There are two dip-positions at angles near about $\theta_{c.m.}$ = $4.8^0$ and within $20-24^0$. It is also
observed that shifting of the dip takes place more for $^{23}$Al at lower angle as compared to $^{22}$Mg. This figure shows clear oscillatory pattern in which magnitude of oscillations are large at smaller angles and keep on decreasing with increase in angles at $E_{Proj.}$= 74 MeV/nucleon.\\

\begin{figure}
\vspace{0.5cm}
\hspace{0.9cm}
\includegraphics[width=1.0\columnwidth,clip=true]{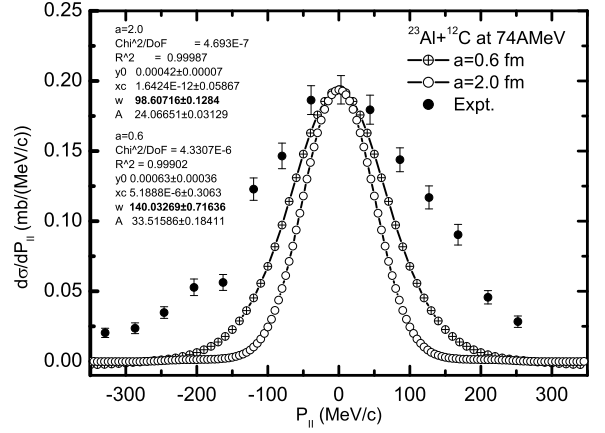}
\label{Fig.1}\vspace{-0.7cm} \caption{Longitudinal momentum distribution of
$^{22}$Mg from the $^{23}$Al+$^{12}$C reaction at projectile energy 74 MeV/nucleon. The experimental data are also given for comparison.}
\end{figure}
%

\subsection {Momentum distribution}
The estimated longitudinal momentum distribution for $^{22}$Mg core of the
reaction $^{23}$Al+$^{12}$C at $E_{Proj}$=74 MeV/nucleon is represented in Fig. 10 along with the experimental 
data \cite{frag2007}. The momentum distributions are estimated by Glauber many body model for diffuseness values of 0.6 and 2.0 fm. The calculated values of momentum distributions underestimate the experimental value. So we have scaled the calculated values to central experimental data point for comparison of both distributions. This figure also signifies that momentum distributions match reasonably with experimental data. The width of this distribution is appeared narrow in comparison to experiment \cite{frag2007} due to large skin. The gaussian fit of full width at half maximum (FWHM) of momentum distribution are found to be 140.02369$\pm 0.72$ MeV/c for 0.6 fm and 98.607 $\pm$ 0.12 MeV/c for 2.0 fm, where as experimental result of Frag et al. \cite{frag2007} is 232$\pm 28$ MeV/c.\\

\section{ Summary and Conclusion}
In summary, the bulk properties like BE, $r_{ch}$ and $\beta_2$ for various
isotopes of Al upto the drip-line regions are studied using RMF(NL3) formalism.
These results are reasonably agreed with experimental data. The one and two proton/neutron separation energy of these nuclei are also explained. Relatively, smaller value of one proton separation energy and other
properties of $^{23}$Al indicates its halo nature. The total nuclear reaction cross-sections $\sigma_R$ for projectiles  $^{23-28}$Al and $^{12}$C as target have been calculated in the
Glauber model using the densities  obtained from RMF(NL3) approximation. The
$\sigma_R$ are in good agreement with the experiments, when we consider the deformation
effect in the densities for isotopic chains. The enhanced reaction cross section of $^{23}$Al at low energy and extended density distribution which seems justified from higher value rms radius as compared to its neighbouring nuclei indicates the existence of halo structure. Higher
value of difference factor `$d$' for $^{23}$Al quantitatively explains its halo
structure. Narrow width in longitudinal momentum distribution also indicates
halo structure of $^{23}$Al. Similar results for $S_{2p}$ as Ref. \cite{wapstra1985} also supports halo nature of $^{23}$Al. More investigations are needed to find specific information regarding the structure of $^{23}$Al.\\

\section{Acknowledgement}
This work is supported by the Department of Science and Technology, Goverment of India, Project No. EMR/2015/002517.

\end{document}